\documentclass[preprint,showkeys,showpacs]{revtex4}

\usepackage{graphics}
\usepackage{graphicx}
\usepackage{amssymb}
\usepackage{amsmath}
\usepackage{amsfonts}
\usepackage{setspace}

\newif\ifFIG
\FIGtrue

\begin{document}

\title{Theoretical study of atoms by the electronic kinetic energy density and stress tensor density}
\author{Hiroo Nozaki}
\author{Kazuhide Ichikawa}
\author{Akitomo Tachibana}
\email{akitomo@scl.kyoto-u.ac.jp}
\affiliation{Department of Micro Engineering, Kyoto University, Kyoto, 615-8540, Japan}

\begin{abstract}
We analyze the electronic structure of atoms in the first, second and third periods using the electronic kinetic energy density and stress tensor density, which are local quantities motivated by quantum field theoretic consideration, specifically the rigged quantum electrodynamics. We compute the zero surfaces of the electronic kinetic energy density, which we call the electronic interfaces, of the atoms. We find that their sizes exhibit clear periodicity and are comparable to the conventional atomic and ionic radii. We also compute the electronic stress tensor density and its divergence, tension density, of the atoms, and discuss how their electronic structures are characterized by them.
\end{abstract}

%\keywords{atoms, kinetic energy density, stress tensor density, tension density, wave function analysis, space partitioning}


%32.10.-f	Properties of atoms
%32.90.+a	Other topics in atomic properties and interactions of atoms with photons (restricted to new topics in section 32)
%31.15.-p	 Calculations and mathematical techniques in atomic and molecular physics (see also 02.70.-c Computational techniques, in mathematical methods in physics)

% 31.10.+z  Theory of electronic structure, electronic transitions, and chemical binding
%31.15.ae Electronic structure and bonding characteristics
%03.65.-w  Quantum mechanics 

\maketitle

%\end{titlepage}

%\setcounter{page}{1}

%\tableofcontents

%%%%%%%%%%%%%%%%%%%%%%%%%%%
\section{Introduction} \label{sec:intro}
%%%%%%%%%%%%%%%%%%%%%%%%%%%
In order to understand the quantum systems from the viewpoint of quantum field theory, in particular quantum electrodynamics (QED), one of the authors has developed
the rigged QED (RQED) theory \cite{Tachibana2001, Tachibana2003, Tachibana2010, Tachibana2013, Tachibana2014a} and our group has applied it to various molecular and condensed matter systems \cite{Tachibana2001,Ikenaga2002, Hotta2002, Hasegawa2003, Yoshida2003, Makita2003, Tachibana2003, Tachibana2004,Kawakami2004, Tachibana2005, Doi2005, Nakamura2005, Nakano2006, Szarek2007, Szarek2008, Fukushima2008, Szarek2009, Ichikawa2009a, Ichikawa2009b,Tachibana2010, Ichikawa2010, Fukushima2010, Ichikawa2011a, Ichikawa2011b, Ichikawa2011c, Fukushima2011, Ikeda2011, Senami2011, Henry2011,Ichikawa2012,Tachibana2013,Tachibana2014a, Nozaki2015}. 
Due to the field theoretic nature, the theory has local quantities defined at each point in space which are useful to describe quantum systems. 
Most important quantities are the electronic kinetic energy density and stress tensor density, whose definitions and meanings are presented in the next section. 
In this section, we shall review our past studies using them. 

As for the electronic kinetic energy density, our definition is proposed in Ref.~\cite{Tachibana2001},  
which is motivated by the quantum field theoretic consideration, and one of its features is that it is not positive-definite.  
Then, it has been proposed that the outermost zero surface of the kinetic energy density gives the intrinsic shape of atoms and molecules. 
This zero surface is designated as ``electronic interface" in Ref.~\cite{Tachibana2001}, and a variety of chemical reactions has been studied using it \cite{Ikenaga2002,Hotta2002,Hasegawa2003,Yoshida2003,Makita2003,Kawakami2004,Tachibana2004,Doi2005,Nakamura2005,Nakano2006,Fukushima2008,Szarek2009,Fukushima2010,Fukushima2011,Ikeda2011,Senami2011,Henry2011,Ichikawa2012}.
We briefly describe some of them below. In Ref.~\cite{Nakamura2005}, the volume surrounded by the outermost electronic interfaces of a cluster model is defined as its ``shape volume" and is used to calculate the electronic dielectric constant of the cluster models of silicate compounds by means of the Clausius-Mossotti equation. In Refs.~\cite{Szarek2009} and \cite{Henry2011}, local reactivity of Pt clusters and Al clusters regarding hydrogen adsorption has been investigated by mapping the regional chemical potential \cite{Tachibana1999,Tachibana2001,Szarek2007,Szarek2008}, which measures the local chemical reactivity, over the electronic interface of the clusters. The same technique is used to study adsorption of lithium atom on carbon nanotube in Ref.~\cite{Senami2011}.

Meanwhile, using the electronic stress tensor density, together with its divergence, which is called the electronic tension density, 
new views on chemical bonding has been proposed \cite{Tachibana2001,Tachibana2003,Tachibana2004,Tachibana2005,Tachibana2010,Tachibana2013,Tachibana2014a}. 
We have shown that the ``Lagrange point", the point between two atoms where the tension density vanishes, can well characterize the chemical bond between them \cite{Szarek2007,Szarek2008}.
We have also shown that the ``Lagrange surface", separatrices of the tension density vector field, may define boundary surfaces of atoms in a molecule \cite{Tachibana2010,Ichikawa2011b, Ichikawa2011c,Tachibana2013,Nozaki2015}. 
It is pointed out that the eigenvalues and corresponding eigenvectors of the electronic stress tensor in the bonding region play important roles for characterizing types of chemical bonding such as covalency \cite{Tachibana2004,Tachibana2005} and metallicity \cite{Szarek2007,Ichikawa2012,Tachibana2013}. In addition, the bonds between some of semimetal atoms are found to have intermediate nature between covalent and metallic bonds in view of the stress tensor density \cite{Nozaki2015}. 

Although we have applied the analyses based on the electronic kinetic energy density and stress tensor density to molecular and periodic systems as above, there have not been a systematic application to atoms. 
It goes without saying that atoms are building blocks for molecules and their electronic kinetic energy density and stress tensor density carry very important information for our analysis. 
Hence, in this paper, we compute the electronic kinetic energy density, stress tensor density, and tension density of the atoms in the first, second and third periods, and investigate how the atoms are characterized by them.

This paper is organized as follows.
In Sec.~\ref{sec:calc}, we show our definition of electronic kinetic energy density and stress tensor density. We also describe our computational setups.
In Sec.~\ref{sec:S}, we investigate the shape of the electronic interfaces of atoms in the first, second and third periods.
In Sec.~\ref{sec:radius}, we study the size of the electronic interfaces and
compare with the conventionally defined atomic and ionic radii.
In Sec.~\ref{sec:intele}, we study the integrated electron density inside the electronic surface. 
In Sec.~\ref{sec:stress}, we show the electronic stress tensor density and tension density of the atoms. 
Finally, Sec.~\ref{sec:conclusion} is devoted to our conclusion.

%%%%%%%%%%%%%%%%%%%%%%%%%%%%%%%%%%%
\section{Theory and calculation methods} \label{sec:calc}
%%%%%%%%%%%%%%%%%%%%%%%%%%%%%%%%%%%

In this paper, we analyze the electronic structure of atoms using quantities such as the electronic stress tensor density and the kinetic energy density. 
They are based on the RQED theory \cite{Tachibana2003} and we briefly describe them in this section. 
For other studies of quantum systems with the stress tensor in somewhat different contexts and definitions, we refer Refs.~\cite{Epstein1975, Bader1980, Bamzai1981a, Nielsen1983, Nielsen1985, Folland1986a, Folland1986b, Godfrey1988, Filippetti2000, Pendas2002, Rogers2002, Morante2006, Tao2008, Ayers2009, Jenkins2011,GuevaraGarcia2011,Finzel2013,Finzel2014a}.
We also refer Refs.~\cite{Ayers2002,GarciaAldea2007,Anderson2010} regarding other definitions of kinetic energy density.

The most basic quantity in RQED is the electronic stress tensor density operator $\hat{\tau}^{\Pi\, kl}_e(x)$  which is defined as \cite{Tachibana2001}.
\begin{eqnarray}
\hat{\tau}^{\Pi\, kl}_e(x) = \frac{i\hbar c}{2} 
\left[
\hat{\bar{\psi}}(x) \gamma^l \hat{D}_{e\, k}(x) \hat{\psi}(x) 
- \left( \hat{D}_{e\, k}(x) \hat{\psi}(x)\right)^\dagger \gamma^0 \gamma^l \hat{\psi}(x)
\right],   \label{eq:op_stress}
\end{eqnarray}
where $\hat{\psi}(x)$ is the four-component Dirac field operator for electrons with the spacetime coordinate $x=(ct, \vec{r})$.
 $c$ denotes the speed of light in vacuum, $\hbar$ the reduced Planck constant, and  $\gamma^\mu$ ($\mu=$0-3) the gamma matrices. 
 The dagger as a superscript is used to express Hermite conjugate, and $\hat{\bar{\psi}}(x) \equiv \hat{\psi}^\dagger(x) \gamma^0$.
The Latin letter indices like $k$ and $l$ express space coordinates from 1 to 3, and repeated indices 
implies a summation over 1 to 3. 
$\hat{D}_{e\, k}(x)$ is the gauge covariant derivative and it is defined as
$\hat{D}_{e\, k}(x) = \partial_k + i\frac{Z_e e}{\hbar c} \hat{A}_k(x)$, where $Z_e = -1$
and
$\hat{A}_k(x)$ is the vector potential of the photon field operator in the Coulomb gauge (${\rm div} \hat{\vec{A}}(x)=0$).
The important property of this quantity is that the time derivative of the electronic kinetic momentum density operator 
$\hat{\vec{\Pi}}_e(x) = \frac{1}{2} \left(
i\hbar \hat{\psi}^\dagger(x) \hat{\vec{D}}_e(x) \hat{\psi}(x) - i\hbar \left( \hat{\vec{D}}_e(x)\hat{\psi}(x) \right)^\dagger \cdot \hat{\psi}(x) \right)$
is expressed by the sum of the Lorentz force density operator $\hat{\vec{L}}_e(x)$ and the tension density operator $\hat{\vec{\tau}}^\Pi_e(x)$,
which is the divergence of $\hat{\tau}^{\Pi\, kl}_e(x)$.
Namely, 
\begin{eqnarray}
\frac{\partial}{\partial t} \hat{\vec{\Pi}}_e(x) &=& \hat{\vec{L}}_e(x) + \hat{\vec{\tau}}^\Pi_e(x),   \label{eq:dPiedt} \\
\hat{\vec{L}}_e(x) &=& \hat{\vec{E}}(x) \hat{\rho}_e(x) + \frac{1}{c} \hat{\vec{j}}_e(x) \times \hat{\vec{B}}(x),  \\
\hat{\tau}^{\Pi k}_e(x) &=& \partial_l \hat{\tau}^{\Pi\, kl}_e(x) \\
&=&
\frac{i\hbar c}{2} \Bigg[ \left(\hat{D}_{el}(x)\hat{\psi}(x) \right)^\dagger \gamma^0 \gamma^l \cdot \hat{D}_{ek}(x) \hat{\psi}(x)
+ \hat{\bar{\psi}}(x) \gamma^l \hat{D}_{ek}(x) \hat{D}_{el}(x) \hat{\psi}(x)  \nonumber \\
& &-\left( \hat{D}_{ek}(x)\hat{D}_{el}(x)\hat{\psi}(x) \right)^\dagger \gamma^0 \gamma^l \cdot \hat{\psi}(x)
-\left( \hat{D}_{ek}(x)\hat{\psi}(x) \right)^\dagger \gamma^0 \gamma^l \cdot \hat{D}_{el}(x) \hat{\psi}(x) \Bigg] \label{eq:op_tension} \nonumber \\
& &
-\frac{1}{c} \left( \hat{\vec{j}}_e(x) \times \vec{B}(x) \right)^k
\end{eqnarray}
where $\hat{\rho}_e(x)$ and $\hat{\vec{j}}_e(x)$ are the electronic charge density operator and charge current density operator respectively, and 
$\hat{\vec{E}}(x)$ and $\hat{\vec{B}}(x)$ denote the electric field operator and magnetic field operator respectively.

As we study nonrelativistic systems in this paper, we approximate the expressions above in the framework of
the primary RQED approximation \cite{Tachibana2013,Tachibana2014a}.
In this approximation, the small components of the four-component electron field are expressed by the large components as $\hat{\psi}_S (x) \approx  - \frac{1}{2m c} i\hbar \sigma^k  D_k \hat{\psi}_L (x)$, where $m$ is the electron mass, and the spin-dependent terms are ignored. Then, we take the expectation value of Eq.~\eqref{eq:dPiedt} with respect to the stationary state of the electrostatic Hamiltonian. This leads to the equilibrium equation as
\begin{eqnarray}
0 = \langle \hat{L}^k_e(x) \rangle + \langle \hat{\tau}^{S k}_e(x) \rangle= \langle \hat{L}^k_e(x) \rangle + \partial_l \langle  \hat{\tau}^{S\, kl}_e(x) \rangle,	\label{eq:L}
\end{eqnarray}
which shows the balance between electromagnetic force and quantum field force at each point in space. Since this expresses the fact that the latter force keeps the electrons in the stationary bound state in atomic and molecular systems, we can study them from the viewpoint of quantum field theory by using the stress tensor density and tension density. We express $\langle \hat{\tau}_e^{S k}(x) \rangle$ and $\langle \hat{\tau}_e^{Skl}(x) \rangle$ respectively $\tau_e^{Sk}(\vec{r})$ and $\tau_e^{Skl}(\vec{r})$ for simplicity. 
Note that, as we consider stationary state, we write only spatial coordinate $\vec{r}$. 
Writing explicitly,
\begin{eqnarray} 
\tau^{Skl}_{e}(\vec{r}) &=& \frac{\hbar^2}{4m}\sum_i \nu_i
\Bigg[\psi^*_i(\vec{r})\frac{\partial^2\psi_i(\vec{r})}{\partial x^k \partial x^l}-\frac{\partial\psi^*_i(\vec{r})}{\partial x^k} \frac{\partial\psi_i(\vec{r})}{\partial x^l} \nonumber\\
& & \hspace{4cm} +\frac{\partial^2 \psi^*_i(\vec{r})}{\partial x^k \partial x^l}\psi_i(\vec{r}) -\frac{\partial \psi^*_i(\vec{r})}{\partial x^l}\frac{\partial \psi_i(\vec{r})}{\partial x^k}\Bigg], \label{eq:stress}
\end{eqnarray}
\begin{eqnarray} 
\tau^{S k}_{e}(\vec{r}) &=&  \partial_l  \tau_e^{Skl}(\vec{r}) \nonumber \\
&=&\frac{\hbar^2}{4m}\sum_i \nu_i
\Bigg[\psi^*_i(\vec{r})\frac{\partial \Delta\psi_i(\vec{r})}{\partial x^k}-\frac{\partial\psi^*_i(\vec{r})}{\partial x^k} \Delta\psi_i(\vec{r}) \nonumber\\
& & \hspace{4cm} +\frac{\partial \Delta\psi^*_i(\vec{r})}{\partial x^k}\psi_i(\vec{r}) -\Delta \psi^*_i(\vec{r}) \frac{\partial \psi_i(\vec{r})}{\partial x^k}\Bigg],
\label{eq:tension}
\end{eqnarray}
where $\psi_i(\vec{r})$ is the $i$th natural orbital and $\nu_i$ is its occupation number.
$\Delta$ denotes the Laplacian, $\Delta \equiv \sum_{k=1}^3 (\partial/\partial x^k)^2$.
The eigenvalue of the symmetric tensor $\stackrel{\leftrightarrow}{\tau}_e^{S}$ is the principal stress and the eigenvector is the principal axis. We denote the eigenvalues as $\tau^{S11}_{e}(\vec{r}) \le \tau^{S22}_{e}(\vec{r}) \le \tau^{S33}_{e}(\vec{r})$.

We note that the tension density in the form of Eq.~\eqref{eq:tension} is same as what is called quantum force density in Refs.~\cite{Pendas2002,Pendas2012}. Then, the Ehrenfest force field used in Refs.~\cite{Pendas2002,Pendas2012,Maza2013,Dillen2015} 
(and the force density in Ref.~\cite{Bader1980}, Eq.~(24)) is same as the tension density with the minus sign in the stationary state. 
Another note is on the ambiguity of the stress tensor density. It is not defined uniquely since mathematically any tensor whose divergence is zero can be added to. Our stress tensor density Eq.~\eqref{eq:stress} is same as the one in Ref.~\cite{Bader1980}, Eq.~(22). The difference in the definitions and approximations are discussed recently in Refs.~\cite{Ayers2009, Jenkins2011, GuevaraGarcia2011,Finzel2013,Finzel2014a}.
We advocate the use of Eq.~\eqref{eq:stress} since it comes from the stress tensor density operator Eq.~\eqref{eq:op_stress}, which is a minimal combination respecting the Lorentz covariance, gauge invariance and hermiticity. Moreover, this definition turns out to be phenomenologically useful as shown by our works mentioned in the previous section.

Another important quantity in the RQED is the electronic kinetic energy density operator defined as
\cite{Tachibana2001},
\begin{eqnarray}
\hat{T}_e(x) &=& -\frac{\hbar^2}{2m}\cdot \frac{1}{2} \left( \hat{\psi}^\dagger(x) \hat{\vec{D}}^2_e(x) \hat{\psi}(x)
+ \left( \hat{\vec{D}}^2_e(x) \hat{\psi}(x) \right)^\dagger \cdot \hat{\psi}(x) \right). \label{eq:op_ked}
\end{eqnarray}
As is done for the electronic stress tensor density operator, 
we apply the primary RQED approximation to Eq.~\eqref{eq:op_ked} and take the expectation value with respect to the stationary state of the electrostatic Hamiltonian.
Then, we obtain the definition for the electronic kinetic energy density as
\begin{eqnarray}
n_{T_{e}}(\vec{r}) = - \frac{\hbar^2}{4m}
\sum_{i} \nu_i \left[ \psi_{i}^{*}(\vec{r}) \Delta \psi_{i}(\vec{r}) + 
\Delta \psi_{i}^{*}(\vec{r}) \cdot \psi_{i}(\vec{r}) \right].   \label{eq:ked} 
\end{eqnarray}
Note that our definition of the electronic kinetic energy density is not positive-definite. 
Using $n_{T_e}(\vec{r})$, we can divide the whole space into three types of region
as follows \cite{Tachibana2001}:
\begin{eqnarray}
R_D &=& \left\{\,\vec{r}\,|\,n_{T_e}(\vec{r}) > 0 \right\} \quad {\rm :\ electronic\ drop\ region}  \\
S &=& \left\{\,\vec{r}\,|\,n_{T_e}(\vec{r}) = 0 \right\} \quad {\rm :\ electronic\ interface}  \\
R_A &=& \left\{\,\vec{r}\,|\,n_{T_e}(\vec{r}) < 0 \right\} \quad {\rm :\ electronic\ atmosphere\ region}  
\end{eqnarray}
In $R_D$, the electronic drop region, the classically allowed motion of electron is guaranteed and the electron density is amply accumulated.
In $R_A$, the electronic atmosphere region, the motion of electron is classically forbidden and the electron density is dried up.
The boundary $S$ between $R_D$ and $R_A$ is called the electronic interface and corresponds to a turning point.
 The outermost $S$ can give a clear image of the intrinsic shape of atoms and molecules and is, therefore, an important region in particular. 
 
Thus, in our method, the outermost electronic interface defines the shape of atoms and molecules. We note that it is frequently defined by the isosurface of the electron density. The values of isosurface like 0.001 a.u. and 0.002 a.u. are proposed in Refs.~\cite{Bader1987,Bader,Popelier}, and there is some arbitrariness. In contrast, since our definition uses the zero isosurface of  non-positive-definite quantity $n_{T_{e}}(\vec{r})$, there is no such arbitrariness. 
It is true that there is ambiguity regarding the definition of the kinetic energy density \cite{Ayers2002,GarciaAldea2007,Anderson2010}, which is sometimes defined as a positive definite quantity. Our definition comes from the electronic kinetic energy density operator, Eq.~\eqref{eq:op_ked}, whose definition is in turn motivated by the relativistic energy dispersion relation between the energy $E$ and momentum $p$: $E = \sqrt{(pc)^2 + (mc^2)^2} \approx mc^2 + \frac{p^2}{2m}$. We take the kinetic energy part $\frac{p^2}{2m}$, replace $p_k$ by $i \hbar \hat{D}_{ek}$ as a usual quantization rule under the existence of the electromagnetic field, and construct a field operator by sandwiching between $\hat{\psi}^\dagger(x)$ and $\hat{\psi}(x)$. We then make the field operator Hermitian as Eq.~\eqref{eq:op_ked} by adding the Hermitian conjugate and divide by two. 
We advocate the use of definition \eqref{eq:ked} since it comes from such a field theoretic construction. 
 
In the end of this section, we summarize our computational setups. The electronic structures used in this paper are obtained by the Gaussian 09 \cite{Gaussian09}.
The computation is performed by the unrestricted Hartree-Fock (UHF) method using the cc-pV5Z basis set \cite{Peterson94}.
To compute the aforementioned quantities 
such as Eqs.~\eqref{eq:stress}, \eqref{eq:tension} and \eqref{eq:ked} from the electronic structure data,
we use the QEDynamics package \cite{QEDynamics} developed in our group.

%%%%%%%%%%%%%%%%%%%%%%%%%%%%%%%
\section{Results and discussion} \label{sec:results}
%%%%%%%%%%%%%%%%%%%%%%%%%%%%%%%

%%%%%%%%%%%%%%%%%%%%%%%%%%%%%%%
\subsection{Shape of the electronic interface of atoms} \label{sec:S}
%%%%%%%%%%%%%%%%%%%%%%%%%%%%%%%

In Fig.~\ref{fig:ei}, we show the electronic interface, $S$, for atoms of elements in the first, second and third periods.
Spin multiplicities are chosen to give their ground states, whose term symbols are shown in Table~\ref{tab:ked} \cite{Moore1949}.
In the case of open-shell systems, the expectation values of $S^2$ are reported in Table~\ref{tab:ked}, showing
that spin contaminations are small. 
As for the electron configuration regarding the $p$ orbitals, we choose them to be occupied  
in the alphabetical order $p_x$, $p_y$, $p_z$.
For instance, the configuration of B is $1s^2 2s^2 2p_x^1$, C is $1s^2 2s^2 2p_x^1 2p_y^1$ and so on. 
In each panel, the atomic nucleus is placed at the origin. 
For all of these elements, the $S$'s have axial symmetry, and some of them have spherical symmetry.
We omit the information of the value of the kinetic energy density in the figure, but we can recognize its sign by
noting that the region in the neighborhood of the nucleus should have positive kinetic energy density, that is to say $R_D$.
Going outward, $R_A$ and $R_D$ appear alternately bounded by $S$ and we should have $R_A$ at infinity. 
In Fig.~\ref{fig:ei_cs}, we show the cross sections of $S$'s to inspect their detailed structures.
Again, in each panel, the atomic nucleus is placed at the origin.
Since we plot to make the symmetry axis coincides with the horizontal axis,
the axes may differ from panel to panel. 
For example, in the panel of B, the horizontal axis corresponds to $x$-axis, but it corresponds to $z$-axis 
in the panel of C.

We can see that the shapes found in Figs.~\ref{fig:ei} and \ref{fig:ei_cs} have the symmetry which is expected from 
the electron configurations of each atom.
Namely, while the $S$'s of H, He, Li, Be, N, Ne, Na, Mg, P and Ar 
are spherically symmetric, those of the other elements, B, C, O, F, Al, Si, S and Cl are only axially symmetric.
In detail, B, O, Al and S are axially symmetric  with respect to the $x$-axis, reflecting the configuration of 
the outermost $p$ subshell: $np_x^1$ for B and Al, and $np_x^2 np_y^1 np_z^1$ for O and S.
As for C, F, Si and Cl, they are axially symmetric with respect to the $z$-axis due to the configuration 
$np_x^1 np_y^1$ for C and Si, and $np_x^2 np_y^2 np_z^1$ for F and Cl.
Also, the periodicity regarding the size of $S$ is manifestly shown: the size increases as the period
increase in the same row, and decreases as the atomic number increases in the same period. 

The common feature for all the elements is that they clearly have outermost $S$ which is homeomorphic to 2-sphere (two-dimensional surface of a ball).
We denote the outermost $S$ as $S_{outer}$.
We next notice that, while H, He and Ne have only one $S$ (only $S_{outer}$), the others have multiple $S$'s.
As for those with multiple $S$'s, most of them have two $S$'s inside $S_{outer}$,
and in addition,
they are both homeomorphic to 2-sphere and the larger $S$ encloses the smaller ones.
When such a structure is found, we denote the innermost $S$ as $S_{inner}$.
In every case, since these two $S$'s are very close to each other, 
it seems that a relatively thin spherical shell region of $R_A$ is formed inside $S_{outer}$.
In other words, the region inside $S_{outer}$ appears to be divided into two $R_D$'s by the 
nearly spherical-shell-shaped $R_A$.
Exceptions are O and F. 
As for O, it has a single $S$ which is homeomorphic to 2-torus (a solid-torus-shaped $R_A$) inside $S_{outer}$. 
As for F, it has two $S$'s inside $S_{outer}$ but they do not form a shell-like region and just two disconnected $R_A$'s
are formed.
(Actually, these features are so fine that it is very hard to see in Fig~\ref{fig:ei}. They can be more easily imagined from
Fig.~\ref{fig:ei_cs} by rotating the cross section with respect to horizontal axis.)
Thus, for both cases, there is only one connected region of $R_D$, and $S_{inner}$ is not formed.
We check whether these features for O and F may change if we use wavefunctions computed by the restricted open-shell HF (ROHF) method.
We find that the ROHF gives visually undistinguishable $S$'s for O and F.
We discuss why they form such $R_A$'s below. 

On looking at these structures, one might associate the pattern of $S$ as the electron shell structure in atoms.
It is tempting to associate $R_D$ which is divided by thin $R_A$ as the shell structure.
The single $R_D$ seen in H and He can be associated with K shell.
From Li to N, they have two $R_D$'s which can be associated with K and L shells.
However, this breaks down at O, F and Ne, which have only one $R_D$ as pointed out above.
Moreover, from Na to Ar, which should have K, L and M shells, they only have two $R_D$'s.
Therefore, $S$ is not a good indicator of the atomic shell structure.
Instead, it may be better to consider two $R_D$'s separated by the thin $R_A$ region as 
the manifestation of ``core" and ``valence" regions in some sense. 
In particular, the core shell can be defined as $R_D$ enclosed by $S_{inner}$ if it exists. 
The irregularities concerning O, F and Ne are interpreted as follows. 
As for O, if there were only three electrons in the 2$p$ subshell as $2p_x^1 2p_y^1 2p_z^1$, the $R_A$ would be
shell-shaped region like the one in N. 
Since there exists one more $2p_x$ electron, the probability of the electron
in the core region is higher so that shell-shaped $R_A$ is deformed to be solid-torus shape as if penetrated by the $2p_x$ orbital. 
Note that the torus's hole is opened in the direction of $x$-axis, which is consistent with the 
direction of $2p_x$ orbital. 
As for F, one more electron in the $2p_y$ orbital is added and it again penetrates in the core region.
Then, the torus-shaped $R_A$ is split in the $y$-direction, and two $R_A$'s remain on the $z$-axis.
Finally, in Ne, the addition of one more electron in the $2p_z$ orbital completely erases these remaining $R_A$'s.

%%%%%%%%%%%%%%%%%%%%%%%%%%%%%%%
\subsection{Size of the electronic interface and comparison with atomic and ionic radii} \label{sec:radius}
%%%%%%%%%%%%%%%%%%%%%%%%%%%%%%%

In this section, we quantify the size of $S_{outer}$ and $S_{inner}$ in the following manner. 
When $S$ is spherical, the size can be readily defined by its radius. 
However, as we have seen in Sec.~\ref{sec:S}, some of the elements have non-spherical $S$.
In that case, we report the longest and shortest distances between nucleus and $S$,
respectively as the upper and lower limits.
Also, as the central value, we report an averaged distance between nucleus and $S$ by computing the radius
of the ball whose volume is equal to the $S$'s volume. 
As for O and F, for which $S_{inner}$ is not defined, we just report the shortest distance between nucleus and $R_A$
inside $S_{outer}$.
The results are summarized in Table \ref{tab:ked} and Fig.~\ref{fig:radius}.
We can confirm the periodicity regarding the size of $S_{outer}$ and $S_{inner}$ in the figure.
We note that, for $S_{inner}$, the deviation from sphere is so small that we cannot recognize the error bars.
As for O and F, the same quantities are computed using the ROHF wavefunctions. 
We find that $S_{outer}$ and $S_{inner}$ of O are respectively
$0.9610^{+0.0446}_{-0.0313}$ and $[0.1533]$, and those of F are respectively $0.8623^{+0.0159}_{-0.0451}$ and $[0.1378]$,
which are almost identical to the UHF results in Table \ref{tab:ked}.

At this stage, it is instructive to examine $S$ of the cations which are constructed by removing all the electrons
in the outermost shell, for example, B$^{3+}$, C$^{4+}$, N$^{5+}$ and so on. 
As a result, we find that all these cations have only one $S$ (only $S_{outer}$) which is spherical, as expected.  
The $S_{outer}$'s radii of the cations are summarized in Table \ref{tab:ked} and Fig.~\ref{fig:radius}.
Then, we see that 
$S_{inner}$ of each atom is almost same as the $S_{outer}$ of the corresponding cation.
This is another reason that we wish to interpret $R_D$ bounded by $S_{inner}$ as a core region.
(In addition, we have observed in our past works that such core regions defined by $S$ are preserved in molecules. 
For example, Figs.~2, 3, 7 and 8 of Ref.~\cite{Ichikawa2012} for Li and Fig.~2 of Ref.~\cite{Nozaki2015} for C.)
As for O and F, although they do not have $S_{inner}$, the sizes of $R_A$ regions inside their $S_{outer}$
are very close to $S_{outer}$ of O$^{+6}$ and F$^{+7}$. 
This supports our interpretation for the irregular shapes of $R_A$ regions in O and F, that they are caused by the penetration of electrons in $2p$ orbitals into their core regions, as argued in the end of Sec.~\ref{sec:S}.

Next, in Fig.~\ref{fig:radius}, we plot atomic radii from Ref.~\cite{Pyykko2009} and ionic radii from Ref.~\cite{Pauling},
along with the size of $S_{outer}$ and $S_{inner}$ as determined above.
We quote two types of ionic radii, crystal radii and univalent radii, which appear in Ref.~\cite{Pauling}.
The crystal radii of multivalent ions are defined so that the sum of two radii is equal to the actual equilibrium
interionic distance in a crystal. 
For example, the univalent radius is applied to Mg$^{2+}$F$_2^-$ while the crystal radius to Mg$^{2+}$O$^{2-}$.
We find that the atomic radii are comparable to the size of $S_{outer}$,
in a sense that they have similar periodicity and differ by a factor of two at most.
Similarly, we can say that the ionic radii are comparable to the size of $S_{inner}$.
However, taking a closer look, the atomic radii are always smaller than the size of $S_{outer}$.
The univalent ionic radii are always larger than the size of $S_{inner}$. 
Although some of the crystal ionic radii are very close to the size of $S_{inner}$, 
since the slopes with respect to the nuclear charge are different, the agreement should be
considered as a mere coincidence. 

Such a disagreement between the atomic radii and $S_{outer}$'s size, and the one between the ionic radii
and $S_{inner}$'s size, are not surprising.
This is because,
while the atomic/ionic radii are defined to give equilibrium internuclear distance in compounds 
and crystals on summing two radii, 
the sizes of $S_{outer}$ and $S_{inner}$ are defined for isolated atoms. 
As for the atomic radii, we can consider that the formation of covalent bond just begins when
the $S_{outer}$'s of two atoms touch each other (this moment is called the ``electronic transition state" \cite{Tachibana2001}). 
Then, after the bond formation, at the equilibrium, 
the interatomic distance is shorter than the sum of the radii of $S_{outer}$ of the two atoms.
Therefore, the formation of chemical bonding makes the atomic radii smaller than the $S_{outer}$'s sizes.
As for the ionic radii, let us consider X$^{n+}$Y$_n^{-}$ for the univalent radii and 
X$^{n+}$Y$^{n-}$ for the crystal radii of multivalent ions.
The simplified description of ionic bonding is that, first, the valence electrons of X
move from X to Y, and then, they are bonded by the electrostatic force between X cation and Y anion. 
Then, the ionic radii of X$^{n+}$ should be about the size of $S_{inner}$ of X.
However, as is well known, this picture is oversimplified and some fraction of the valence electrons
remain at X. This is the reason why the univalent radii are larger than the $S_{inner}$'s sizes. 
Similar consideration applies for the crystal radii of multivalent ions, but in this case, 
the electrostatic attraction is so strong that the interionic distance becomes much shorter.
This is the reason why the crystal radii are smaller than the $S_{inner}$'s sizes for large $n$. 

%%%%%%%%%%%%%%%%%%%%%%%%%%%%%%%
\subsection{Integrated electron density inside the electronic interface} \label{sec:intele}
%%%%%%%%%%%%%%%%%%%%%%%%%%%%%%%

In this section, we study $S$ from the viewpoint of amount of electrons enclosed within it.
As is mentioned in Sec.~\ref{sec:calc}, $R_D$ is the region where classically allowed motion of electron is guaranteed
and the electron density is amply accumulated.
We demonstrate this point by integrating electron density over $R_D$ in each atom. 
In other words, we integrate over the region inside $S_{outer}$ omitting $R_A$, if it exists inside $S_{outer}$ like the cases of O and F. 
The results are shown in Fig.~\ref{fig:intele}, where the upper panel shows the integrated electron density and
the lower panel shows its ratio to the total electron number. 
We can confirm that electrons have high probability, about 70\%--90\%, to be found in $R_D$.
The results for the cations are also shown and we find that the ratios fall within the similar range. 

We note here on the analytic result for the ground state of the hydrogen-like atom with nuclear charge $Z$.
When we use the analytic wave function, $R_D$ is found to be a ball whose radius is $2/Z$\,bohr \cite{Tachibana2001}, and
the integration of electronic density over there is $1 - 13/e^4 \approx 0.761897$, where $e$ is
the Euler's number. This integration does not depend of $Z$.
Our numerical integration for H reproduces this value with the error less than 1\%.

In Fig.~\ref{fig:intele}, upper panel, and in Table \ref{tab:compare},
we also show the results of integration only over the core region defined as the region enclosed by $S_{inner}$. For the atoms which do not exhibit $S_{inner}$ (H, He, O, F and Ne), data points are omitted. 
The results are almost same as those over the region within $S_{outer}$ of corresponding cations, which are expected from the closeness of their sizes as seen in the previous section. 
The integrated electron density of core region turns out to be about 1.5--1.7 for the atoms of the second period and about 8.9--9.4 for those of the third period. 
In Table \ref{tab:compare}, we also quote core radius and core electron number derived from some of the shell structure descriptors in the literature, Ref.~\cite{Schmider1992} considering ``ideal shells", Ref.~\cite{Kohout1996} using the electron localization function, and Ref.~\cite{Wagner2011} based on inhomogeneity measures of the electron density in connection with the electron localizability indicator. Our values based on $S_{inner}$ are generally much smaller than the others, indicating that the core region defined by $S_{inner}$ has somewhat different meaning from the one based on the shell structure descriptors.
 Although our core electron numbers are away from %2 and 10 respectively of 
the values of the ideal shells \cite{Schmider1992}, the relation between core-valence partitioning in the orbital picture and one in real space is not straightforward and whether a partitioning scheme has to be an ideal one is not evident \cite{Finzel2014b}. 
We believe that our definition of a core region using the electronic interface is worth investigating further, along with other definitions in the literature \cite{Politzer1976,Boyd1977,Politzer1977,Smith1977,Kohout1991,Sen1993,Kakkar1994,Sen1995,Desmarais1996,Kohout1996,Sen1996,Kohout1997,Sahni2001,Navarrete2008,Naka2010,Wagner2011}, for describing chemical bonding and reactions.

%%%%%%%%%%%%%%%%%%%%%%%%%%%%%%%%%%%%%%%%%%%%%%%
\subsection{Stress tensor density and tension density} \label{sec:stress}
%%%%%%%%%%%%%%%%%%%%%%%%%%%%%%%%%%%%%%%%%%%%%%%
In this section, we examine the electronic stress tensor density and tension density of the atoms. 
We start from the tension density, which is shown in Fig.~\ref{fig:tension} as black arrows. 
They are normalized and their norm is expressed by the color map. 
The electronic interfaces are plotted by the black solid lines, which are same as Fig.~\ref{fig:ei_cs}.
We see that, for all the atoms we examine, the arrows go radially from the center, at which the nucleus is located. 
There are some exceptions but they are found at very far outside $S_{outer}$ so that we may regard them as numerical artifacts. 
Such artifacts have been discussed in Ref.~\cite{Ichikawa2010} by comparing the tension density computed from the exact solution for H$_2^+$ with one from solutions using gaussian basis sets. Recently, Ref.~\cite{Dillen2015} has examined the artifacts of the Ehrenfest force density (which is the minus of the tension density) based on the Slater-type orbitals.
This pattern of the atomic tension density that the vector field points from the nucleus to outside of the atom is important for the formation of ``Lagrange surface" of a molecule \cite{Tachibana2010,Ichikawa2011b, Ichikawa2011c,Tachibana2013,Nozaki2015}. 
The Lagrange surface of a molecule AB composed of atoms A and B is defined as follows \cite{Tachibana2010,Tachibana2013}.
The Heisenberg uncertainty principle allows electron to diffuse away from each atomic center it belongs, and
the diffusion force is the tension compensating the Lorentz force exerting from each atomic center as expressed by Eq.~\eqref{eq:L}.
The tension vector fields originated from A and B mutually collide to form a separatrix that discriminates each region of atomic center.
The separatrix is called the Lagrange surface. We refer Fig.~2 of Ref.~\cite{Tachibana2010} or Fig.~12.3 of Ref.~\cite{Tachibana2013} for the schematic picture of the Lagrange surface, and Fig.~7 of Ref.~\cite{Nozaki2015} for a computed example.
This picture of local equilibrium in the stationary state of a molecule assumes that the tension of an isolated atom has a pattern such that the vectors radiate from the nucleus. Although there is no mathematical proof that all the atoms exhibits the pattern, our calculation shows that the pattern holds for those in the first, second and third periods.

Here, some comments on the Lagrange surface may be in order. In our method, the boundary between atoms are defined by this Lagrange surface (the outer boundary is defined by the outermost electronic interface as described in Sec.~\ref{sec:calc}). Although this is conceptually similar to the so-called interatomic surface \cite{Bader,Popelier}, the definitions are different. The interatomic surface is defined as a separatrix of the gradient vector field of the electron density whereas the Lagrange surface is one of the tension vector field. 
As is noted in Sec.~\ref{sec:calc}, the tension density and the Ehrenfest force density is only different by a minus sign so that separatrices in these vector fields coincide. Therefore, the recent work in Ref.~\cite{Pendas2012} on the partitioning scheme based on separatrices of the Ehrenfest force field is same as one using the tension density and Lagrange surface in our terminology.

Next, we move on to study the stress tensor density. 
In Fig.~\ref{fig:stress}, the largest eigenvalue of the stress tensor $\tau_e^{S33}(\vec{r})$ is plotted as a color map and corresponding eigenvector is expressed by a black rod. One may notice that there are some regions in B, Al and S where no rod is shown. There, the eigenvectors are directed perpendicular to the plane so that a rod becomes a dot and not shown. Also, one may notice that some rods are made thicker and colored orange, for example in Li and Be. At these points, the largest two eigenvalues, $\tau_e^{S33}(\vec{r})$ and $\tau_e^{S22}(\vec{r})$, are degenerate. For the atoms we have computed here, one of the two eigenvectors corresponding to the degenerated eigenvalue is always in the direction perpendicular to the plane so that only one (thick orange) rod appears at each point. 
In Fig.~\ref{fig:stress}, the zero isosurfaces of the eigenvalue are plotted by the red solid lines and the electronic interfaces are plotted by the black solid lines. %Note the different use of line types from previous figures. 
As for the zero isosurfaces of the eigenvalue, we note that some of them appear as numerical artifacts.
This is most conspicuously shown in H. Although the analytic wave function of the H atom gives $\tau_e^{S33}(\vec{r})=0$ everywhere \cite{Tachibana2004}, we see several zero-surfaces in the figure of H. They are numerical artifacts caused by the oscillatory behavior around zero, which is in turn due to the approximate wave function constructed from gaussian functions. 
Then, it is not hard to imagine similar artificial zero-surfaces may appear when $\tau_e^{S33}(\vec{r})$ is close to zero, typically far away from the nucleus.
Although we cannot tell which zero-surface is the artifact a priori, we may regard the one which locates well outside $S_{outer}$ as artificial. 

Before analyzing the stress tensor of the atoms further, it may be instructive here to summarize our past findings on the stress tensor of molecules some more in detail than Sec.~\ref{sec:intro}.
 In Ref.~\cite{Tachibana2004}, it has been proposed that the bonding region with covalency can be characterized and visualized by the ``spindle structure", where the largest eigenvalue of the electronic stress tensor is positive (tensile stress) and the corresponding eigenvectors form a bundle of flow lines that connects nuclei. 
 In other words, the positive largest eigenvalue can be associated with the Lewis electron pair formation. 
In contrast, the eigenvalue can be negative (compressive stress) at bonding region for several cases. For one thing, it is found at a very short bond such as C$_2$H$_2$ \cite{Tachibana2005,Szarek2008,Ichikawa2011b}.  
It is explained that the internuclear region of C$_2$H$_2$, which should be covalent, is overwhelmed by the atomic compressive stress around the C nuclei as they are so close. The atomic compressive stress in turn is attributed to marginal stability around atoms \cite{Tachibana2005}.
Another case is when a bond exhibits metallicity \cite{Szarek2007,Ichikawa2012,Tachibana2013,Nozaki2015}. In this case, the eigenvalues are not only negative but also degenerate, $0>\tau_e^{S33}(\vec{r}) \approx \tau_e^{S22}(\vec{r}) \approx \tau_e^{S11}(\vec{r})$, reflecting the liquid-like (compressive and isotropic) nature of metallicity of chemical bonding. (In passing, for the case of  C$_2$H$_2$, the eigenvalues are not degenerate, $0 > \tau_e^{S33}(\vec{r}) >  \tau_e^{S22}(\vec{r}) \approx \tau_e^{S11}(\vec{r})$ \cite{Szarek2007,Ichikawa2012}.)
 
With these in mind, we will now take a look at Fig.~\ref{fig:stress}. 
An overall rough trend is that the eigenvalue is negative, manifesting the atomic compressive stress \cite{Tachibana2005}. There are, however, some places with positive eigenvalues (we omit the discussion for H, whose largest eigenvalue is zero everywhere as mentioned above, as it has already been discussed in our previous works \cite{Tachibana2001, Tachibana2004,Tachibana2005} ). We see the positive eigenvalue regions in: (i) entire $R_D$ of He, (ii) central regions of the atoms in the second period, and (iii) outer regions of $R_D$ including some parts or all of $S_{outer}$ in O, F, and Ne. 
 The case (i) is interpreted as due to the electron pairing of  $1s^2$.
The case (ii) is considered to have the same origin as the case (i), though the positive regions shrink as outer shells pile up. As we go to the third period, such positive regions disappear from the core regions. It seems that the tensile stress caused by the electron pairing is immersed under the atomic compressive stress. The case (iii) is considered to occur due to the electron pairing in the $2p$ shell. As for O, the electron pairing of $2p_x^2$ is manifested as the positive eigenvalue regions, so that they locate in the $x$-directions. Similarly, the electron pairing of $2p_x^2$ and $2p_y^2$ in F leads to the positive eigenvalue region axially symmetric with respect to the $z$-axis, and that of $2p_x^2$, $2p_y^2$, and $2p_z^2$ in Ne results in the spherically symmetric positive eigenvalue region. Again, it seems that such tensile stress is immersed under the atomic compressive one in the third period atoms. 

Another noticeable feature, such as found in Li, is that the valence region has  two layers.
The outside layer consists of the eigenvectors going radially, and the inside layer is characterized by the two degenerate eigenvalues whose eigenvectors are perpendicular to the radial direction. 
The degenerate character of the inside layer has been found to appear in a Li$_2$ molecule or other small Li clusters \cite{Ichikawa2012}. We have pointed out in Ref.~\cite{Ichikawa2012} that, in view of the stress tensor density, chemical bonding of the Li clusters, even Li$_2$, is characterized by a lack of directionality which is represented by the degenerate eigenvalues and corresponding eigenvectors. The stress tensor density of the Li atom is found to already exhibit such a degenerate feature. 
We see in Fig.~\ref{fig:stress} that Be, Na, Mg, P, and Ar exhibit similar pattern to Li. 
How these degenerate features may play a role in compounds and clusters involving these atoms 
is interesting question to ask, and this is work in progress in our group.

%%%%%%%%%%%%%%%%%%%%%%%%%%%%
\section{Conclusion} \label{sec:conclusion}
%%%%%%%%%%%%%%%%%%%%%%%%%%%%

In this paper, we have analyzed the electronic structure of atoms in the first, second and third periods
using the electronic kinetic energy density, tension density, and stress tensor density, 
which are local quantities motivated by quantum field theoretic consideration, specifically RQED.

We have used the concept of the electronic interface $S$, which has been defined in Ref.~\cite{Tachibana2001} 
as the zero surface of non-positive-definite kinetic energy density to illustrate the shape of atoms and molecules.
We have investigated the shape of $S$ and found that $S_{outer}$, the outermost $S$, exhibits the symmetry which is expected from the electron configurations of each atom.
We have pointed out that, in most cases, the region inside $S_{outer}$ is divided into two $R_D$'s by the 
nearly spherical-shell-shaped $R_A$. 
For these two $R_D$'s, we have associated the inner $R_D$ as the core region and the outer $R_D$ as the valence region.
This interpretation is supported by the size of $S_{outer}$ of the cations which are constructed by removing all the electrons
in the outermost shell.
We have denoted the boundary surface of core shell as $S_{inner}$, and have investigated the size of $S_{outer}$ and $S_{inner}$.
Then, we have found the clear periodicity for the size of $S_{outer}$ and $S_{inner}$ with respect to the nuclear charge. 
The size of $S_{outer}$ and $S_{inner}$ have been compared respectively with the atomic and ionic radii in the literature,
and their difference has been discussed. 

As for the tension density, all of the atoms we have examined exhibit a common pattern that the vector field points from the nucleus to outside of the atom.
Such radial pattern of the atomic tension density is important to be confirmed, as it is necessary for the universal local equilibrium picture of electronic stationary state of a molecule with the Lagrange surface \cite{Tachibana2010,Tachibana2013}.
As for the stress tensor density, we have studied its largest eigenvalue and corresponding eigenvector, which has played an important role in our chemical bonding and reactivity analyses on molecular systems. 
We have first confirmed that the atoms generally exhibit negative eigenvalues, the compressive stress, by which marginal stability around atoms is represented \cite{Tachibana2005}. We have also found some regions with positive eigenvalue, the tensile stress, for some atoms.
They can be interpreted as the tensile stress associated with the covalency which describes the electron pairing \cite{Tachibana2004,Tachibana2005}.

We believe that this work has revealed novel aspects of electronic structure of atoms in view of the electronic kinetic energy density and stress tensor density. At the same time, it raises many interesting issues that will need further research. 
First of all, note that our interpretation here is based on the real-valued wave functions obtained by the HF method. Configuration interaction method should be used to respect appropriate symmetry of wave functions, which will be used to study the electronic interfaces and stress tensor in our future work. 
It is also necessary to study the atoms in higher periods to confirm the interpretation presented in this study and to investigate such effects as
the lanthanide contraction and contraction of the heavy elements by relativistic effects.
In addition to the electronic kinetic energy density and stress tensor density, one of the authors has proposed novel quantities related to the electronic spin: spin torque density, zeta force density and spin vorticity \cite{Tachibana2010,Hara2012,Tachibana2012,Tachibana2013,Tachibana2014a,Tachibana2014b,Tachibana2015}. 
The study through these quantities would give us further new insight into our understanding of electronic structure of atoms, and chemistry.

\noindent 
%%%%%%%%%%%%%%%%%%%%%%%%%%%%%%%%%%%%
\section*{Acknowledgment}
%%%%%%%%%%%%%%%%%%%%%%%%%%%%%%%%%%%%
Theoretical calculations were partly performed using Research Center for Computational Science, Okazaki, Japan.
This work was supported by JSPS KAKENHI Grant Number 25410012 and 26810004.
H.~N. is supported by the Sasakawa Scientific Research Grant from the Japan Science Society.

%%%%%%%%%%%%%%%%%REFERENCES%%%%%%%%%%%%%%%%%%%%%%%%%

%%%%%%%%%%%%%%%%%%%%%%%%%%%%%%%%%%%%%%%%%%%%%%%%%%%%%%%%%%%%%%%

\newpage

\begin{table}
\begin{center}
\begin{tabular}{|c|c|c|c|c|c|c|c|c|c|}
\hline
&\multicolumn{5}{|c|}{Neutral atom}
&\multicolumn{2}{|c|}{Cation}
&\multicolumn{2}{|c|}{Radius}\\
\hline
& State & $\langle S^2 \rangle $  & Ratio&$S_{outer}$&$S_{inner}$ &Ratio&$S_{outer}$ &Covalent&Ionic\\
\hline
H & $^2$S$_{1/2}$	&0.750 &0.767&1.067&---&---&---&0.32&---\\
He &  $^1$S$_{0}$	& ---  &0.731&0.623&---&---&---&0.46&---\\

\hline
Li &$^2$S$_{1/2}$	&0.750  &0.740&2.6953&0.3915&0.741&0.3910&1.33&0.60(0.60)\\
Be &$^1$S$_{0}$		&---    &0.744&1.7901&0.2866&0.748&0.2858&1.02&0.31(0.44)\\
B &$^2$P$_{1/2}$	&0.761  &0.782&$1.4855^{+0.2250}_{-0.1630}$&$0.2304^{+0.0066}_{-0.0039}$
&0.753&0.2259&  0.85&0.20(0.35)\\
C&$^3$P$_{0}$		& 2.010 &0.811&$1.2530^{+0.0640}_{-0.1820}$&$0.1955^{+0.0041}_{-0.0086}$
&0.755&0.1862&  0.75&0.15(0.29)\\
N&$^4$S$_{3/2}$		& 3.758 &0.839&1.0828&0.1730&0.756&0.1584&0.71&0.11(0.25)\\
O&$^3$P$_{2}$		& 2.009 &0.871&$0.9602^{+0.0539}_{-0.0333}$&[0.1535]
&0.756&0.1377&  0.63&0.09(0.22)\\
F& $^2$P$_{3/2}$	& 0.754 &0.887&$0.8623^{+0.0173}_{-0.0445}$&[0.1379]
&0.757&0.1219&  0.64&0.07(0.19)\\
Ne& $^1$S$_{0}$		& ---   &0.889&0.7804&---&0.758&0.1093&0.67&---\\

\hline
Na& $^2$S$_{1/2}$	& 0.750 &0.870&2.9140&0.6521&0.893&0.6515&1.55&0.95(0.95)\\
Mg& $^1$S$_{0}$		& ---   &0.857&2.1848&0.5587&0.893&0.5569&1.39&0.65(0.82)\\
Al& $^2$P$_{1/2}$	& 0.771 &0.868&$2.0520^{+0.4548}_{-0.3297}$&$0.4959^{+0.0000}_{-0.0015}$
&0.896&0.4922&1.26&0.50(0.72)\\
Si&$^3$P$_{0}$		& 2.016 &0.875&$1.8148^{+0.1142}_{-0.3351}$&$0.4446^{+0.0002}_{-0.0008}$
&0.897&0.4396&1.16&0.41(0.65)\\
P&$^4$S$_{3/2}$		& 3.751 &0.879&1.6028&0.4032&0.898&0.3972&1.11&0.34(0.59)\\
S&$^3$P$_{2}$		& 2.013 &0.884&$1.4444^{+0.0458}_{-0.0222}$&$0.3706^{+0.0005}_{-0.0009}$
&0.898&0.3620&1.03&0.29(0.53)\\
Cl& $^2$P$_{3/2}$	& 0.761 &0.890&$1.3107^{+0.0115}_{-0.0214}$&$0.3440^{+0.0003}_{-0.0020}$
&0.898&0.3329&0.99&0.26(0.49)\\
Ar& $^1$S$_{0}$		& ---   &0.898&1.1955&0.3221&0.899&0.3082&0.96&---\\

\hline

\end{tabular}
\end{center}
\caption{The ratio refers to the quotient of electron density integrated over $R_D$ inside $S_{outer}$ divided by the total electron number.
Numbers in the columns of $S_{outer}$ and $S_{inner}$ are the measures of their size which are described in Sec.~\ref{sec:radius}, in units of \AA.
Covalent radii are taken from Ref.~\cite{Pyykko2009} and ionic radii from Ref.~\cite{Pauling}, also in units of \AA.
Two types of ionic radii, the crystal radii and univalent radii, are quoted, and the latter is denoted in the parentheses. }

\label{tab:ked}

\end{table}

\begin{table}
\begin{center}
\begin{tabular}{|c|c|c|c|c|c|c|c|c|}
\hline
&\multicolumn{4}{|c|}{Core radius}
&\multicolumn{4}{|c|}{Core electron number} \\
\hline
& $S_{inner}$ & \cite{Schmider1992}  &  \cite{Kohout1996} & \cite{Wagner2011}  & $S_{inner}$ & \cite{Schmider1992}  &  \cite{Kohout1996} & \cite{Wagner2011} \\
\hline
Li	&	0.3915		&	0.81073	&	0.810	&	0.6949	&	1.48	&	2	&	2.0	&	1.9	\\
Be	&	0.2866		&	0.52136	&	0.540	&	0.4316	&	1.53	&	2	&	2.0	&	1.9	\\
B	&	0.2304		&	0.37144	&	0.40	&	0.3082	&	1.57	&	2	&	2.0	&	1.9	\\
C	&	0.1955		&	0.28424	&	0.31	&	0.2347	&	1.62	&	2	&	2.1	&	1.8	\\
N	&	0.1730		&	0.22855	&	0.25	&	0.1869	&	1.69	&	2	&	2.1	&	1.8	\\
O	&	[0.1535]	&	0.19065	&	0.21	&	0.1542	&	---	&	2	&	2.1	&	1.7	\\
F	&	[0.1379]	&	0.16285	&	0.18	&	0.1301	&	---	&	2	&	2.1	&	1.7	\\
Ne	&	---		&	0.14172	&	0.16	&	0.1116	&	---	&	2	&	2.1	&	1.7	\\
\hline
Na	&	0.6521		&	1.12544	&	1.130	&	1.0123	&	8.92	&	10	&	10.1	&	9.9	\\
Mg	&	0.5587		&	0.86178	&	0.892	&	0.7484	&	8.96	&	10	&	10.1	&	9.8	\\
Al	&	0.4959		&	0.71646	&	0.738	&	0.6152	&	9.03	&	10	&	10.1	&	9.7	\\
Si	&	0.4446		&	0.61204	&	0.626	&	0.5209	&	9.08	&	10	&	10.1	&	9.7	\\
P	&	0.4032		&	0.53293	&	0.542	&	0.4509	&	9.14	&	10	&	10.1	&	9.5	\\
S	&	0.3706		&	0.47150	&	0.483	&	0.3974	&	9.22	&	10	&	10.1	&	9.5	\\
Cl	&	0.3440		&	0.42206	&	0.432	&	0.3546	&	9.29	&	10	&	10.1	&	9.4	\\
Ar	&	0.3221		&	0.38145	&	0.391	&	0.3199	&	9.39	&	10	&	10.1	&	9.4	\\\hline

\end{tabular}
\end{center}
\caption{Core radius and core electron number derived from $S_{inner}$ and those from the shell structure descriptors proposed in Refs.~\cite{Schmider1992,Kohout1996,Wagner2011} (the core electron numbers of Ref.~\cite{Schmider1992} are 2 for the atoms of the second period and 10 for those of the third period by definition). 
The core radius is in units of \AA. The core radius defined by $S_{inner}$ is quoted from Table \ref{tab:ked}. %The core electron number is the integrated  electron density over the region defined by the core radius. 
}

\label{tab:compare}

\end{table}

%%%%%%%%%%%%%%%%%%%%%%%%%%%%%%%%%%%%%%%%%%%%%%%%%%%%%%%%%%%%%%%
\ifFIG

\clearpage

\begin{figure}
\begin{center}
\includegraphics[angle=90,width=9cm]{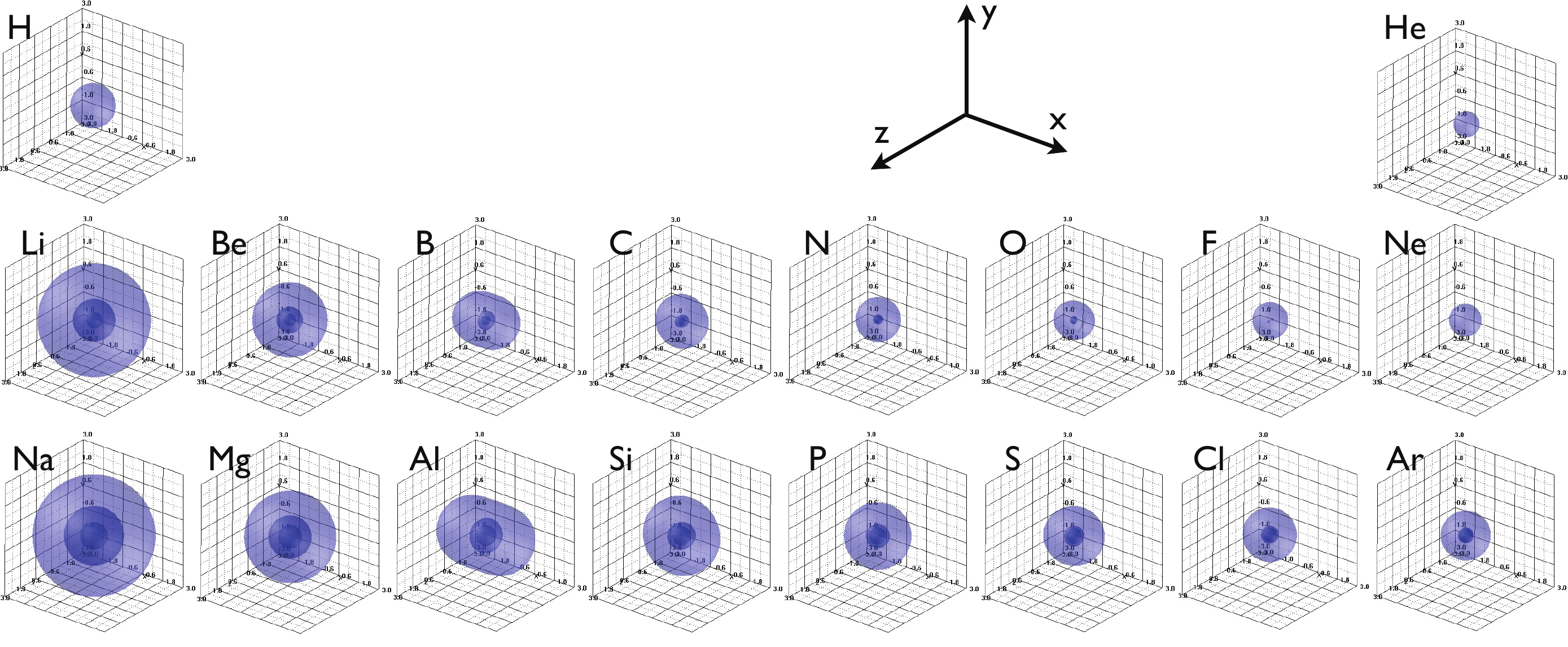}
\caption{Electronic interfaces of the atoms. The units are in \AA.
}		
\label{fig:ei}
\end{center}
\end{figure}

\begin{figure}
\begin{center}
\includegraphics[width=14cm]{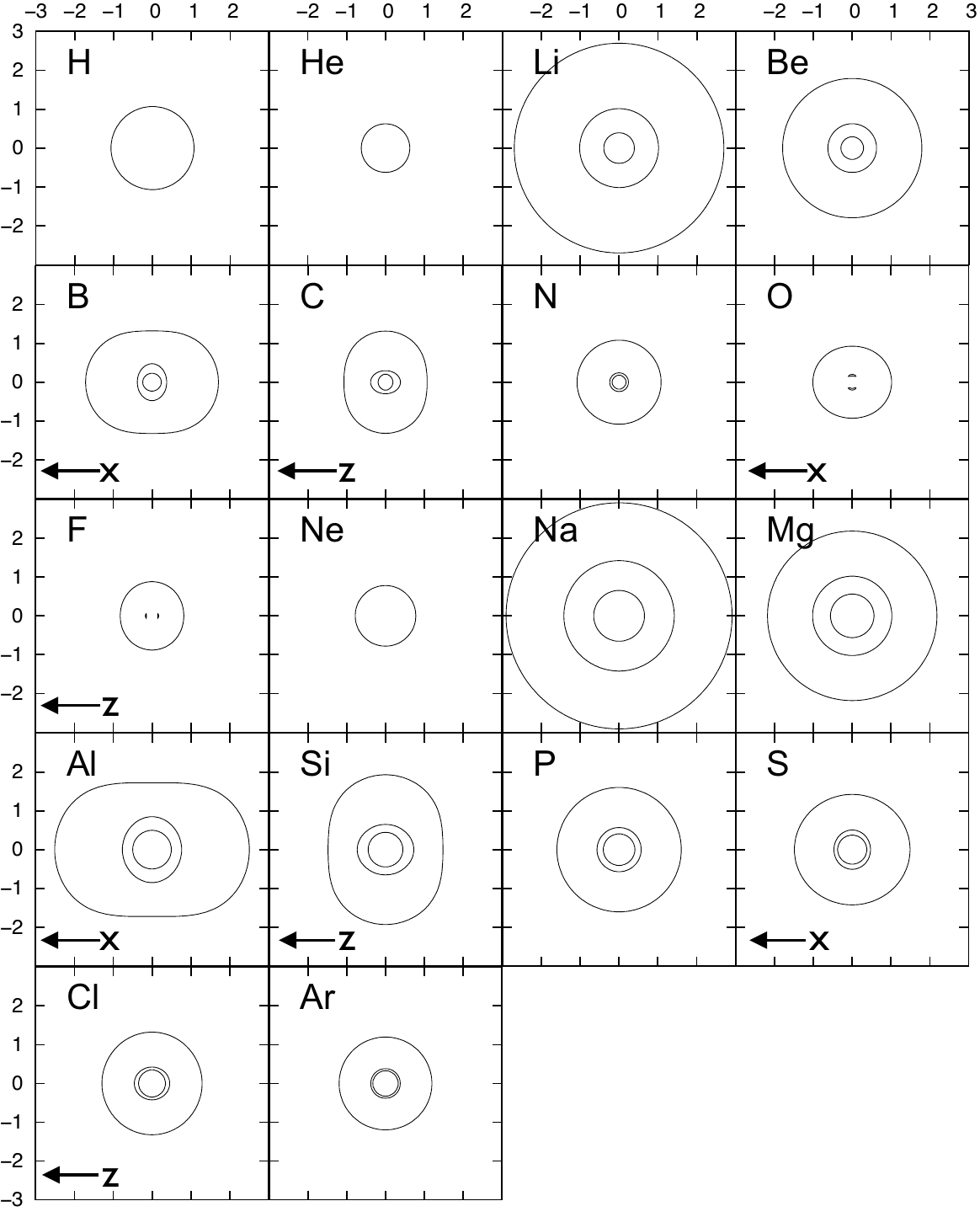}
\caption{Cross sections of the electronic interfaces of the atoms. 
The horizontal axes coincide with the axial symmetry axes of the electronic interfaces. 
The units are in \AA.
}		
\label{fig:ei_cs}
\end{center}
\end{figure}

\begin{figure}
\begin{center}
\includegraphics[width=14cm]{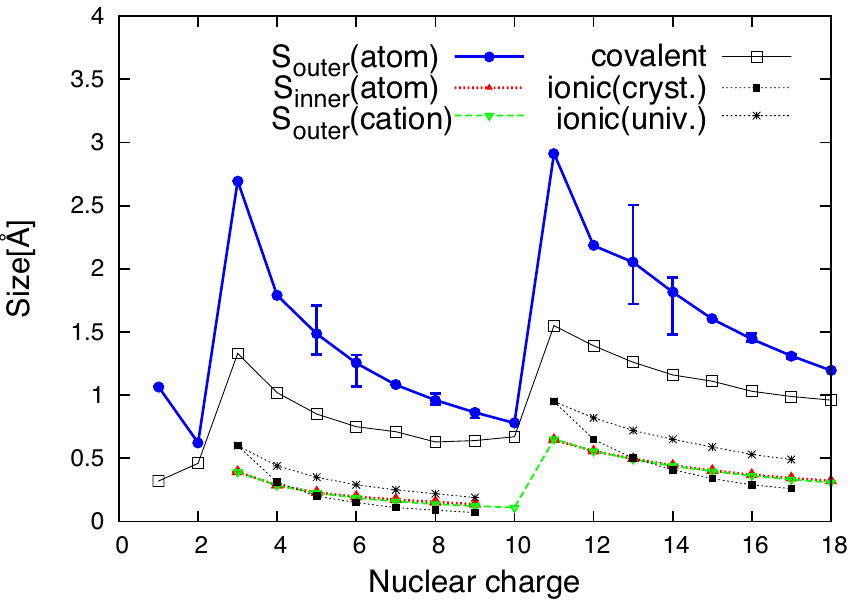}
\caption{The size of the electronic interfaces, which is defined in Sec.~\ref{sec:radius}, are plotted
for $S_{outer}$ (blue solid line) and $S_{inner}$ (red dotted line) of atoms, and $S_{outer}$ of cations (green dashed line).
Covalent radii from Ref.~\cite{Pyykko2009} (black thin solid line) and ionic radii from Ref.~\cite{Pauling} (black thin dotted line) 
are plotted, too. 
As for the ionic radii, the crystal radii (line with filled-square) and univalent radii (line with asterisk) are plotted. 
}		
\label{fig:radius}
\end{center}
\end{figure}

\begin{figure}
\begin{center}
\includegraphics[width=14cm]{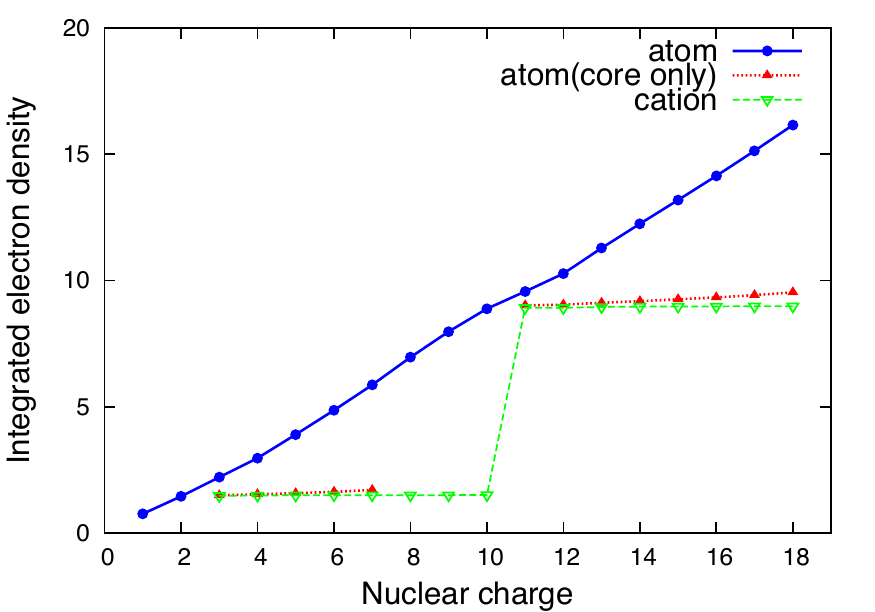}
\includegraphics[width=14cm]{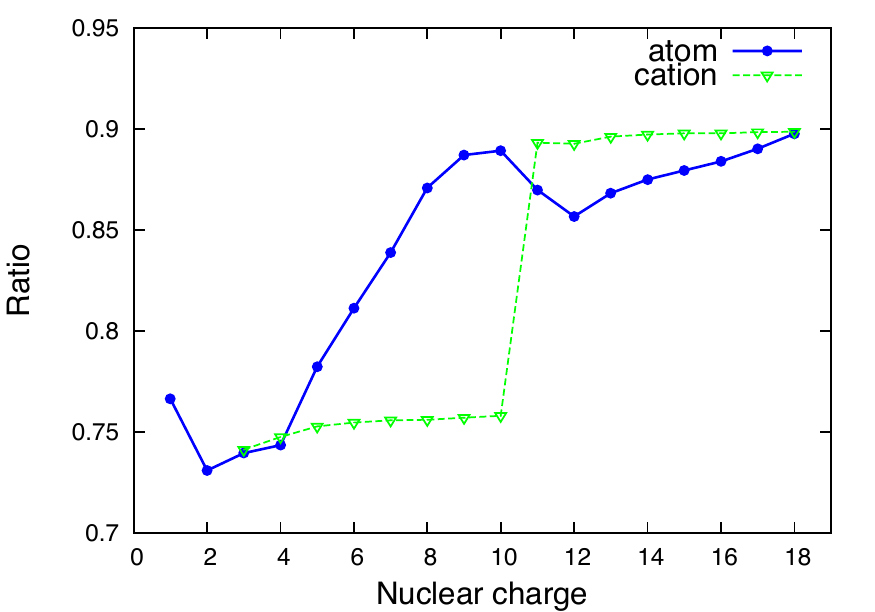}
\caption{In the upper panel, the electron density integrated over the total $R_D$ of atoms (blue solid line), 
$R_D$ of core region of atoms (red dotted line), and total $R_D$ of cations (green dashed line), are plotted.
In the lower panel, the ratios of the electron density integrated over $R_D$ to the total electron number are 
plotted for atoms (blue solid line) and cations (green dashed line).
}		
\label{fig:intele}
\end{center}
\end{figure}

\begin{figure}
\begin{center}
\includegraphics[angle=90,width=8.5cm]{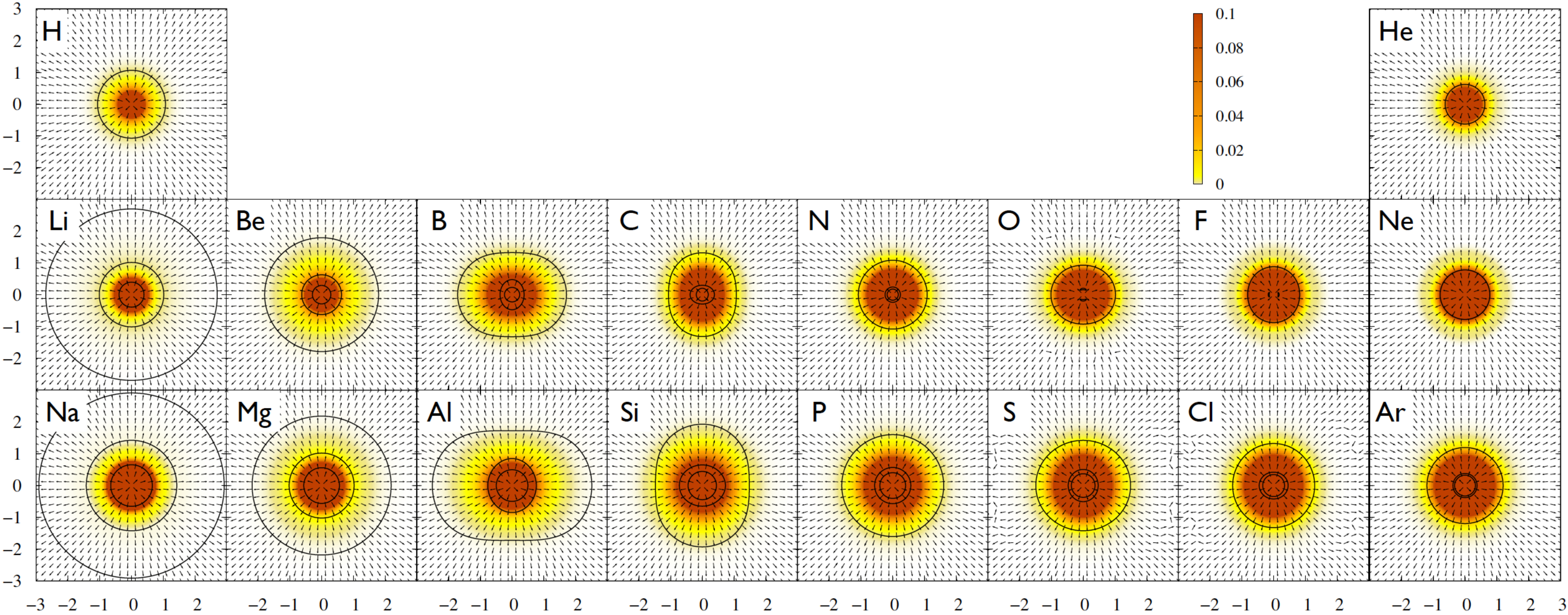}
\caption{The normalized tension density is plotted by the black arrows. Its norm is expressed by the color map. The black solid lines are the electronic interfaces as in Fig.~\ref{fig:ei_cs}.
}		
\label{fig:tension}
\end{center}
\end{figure}

\begin{figure}
\begin{center}
\includegraphics[angle=90,width=8.5cm]{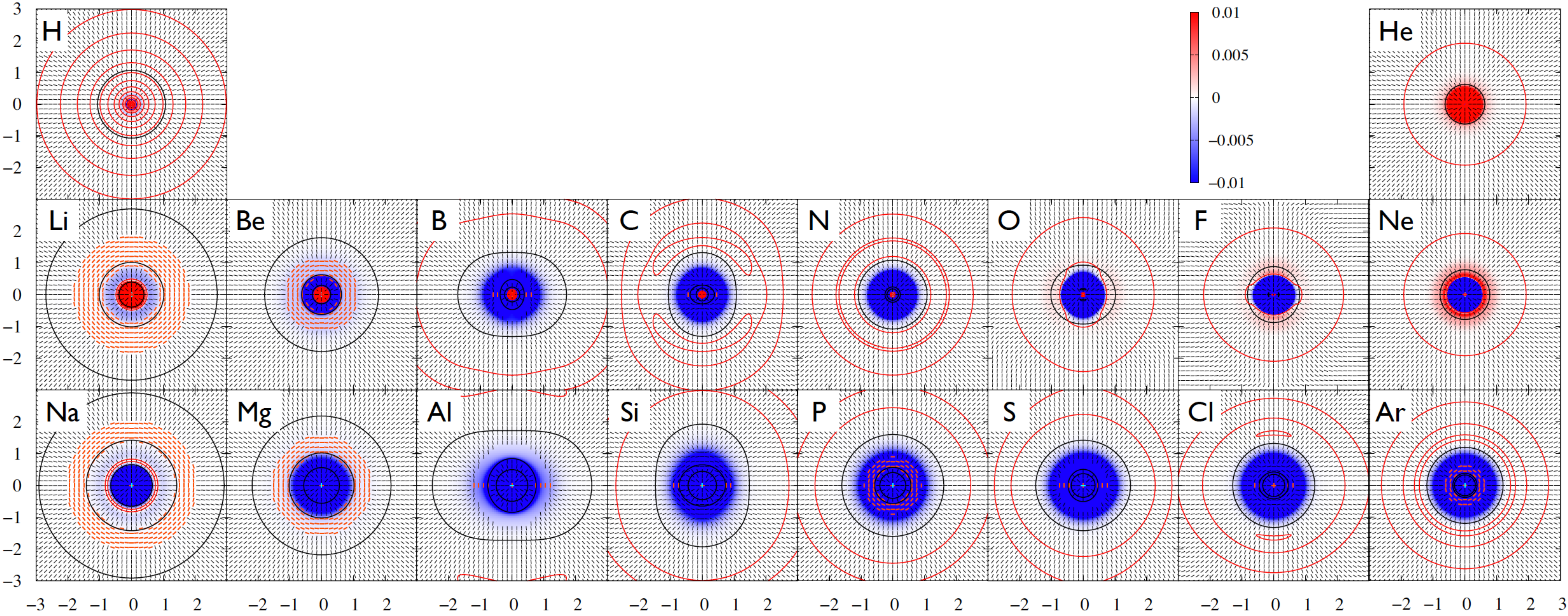}
\caption{The largest eigenvalue of the electronic stress tensor density (color map)
and corresponding eigenvector (black rods) are plotted. 
The thick orange rods imply that the largest two eigenvalues are degenerated (see details for the text).
Note that the red solid lines express the zero isosurfaces of the eigenvalue.
The black solid lines are the electronic interfaces as in Fig.~\ref{fig:ei_cs}.
}		
\label{fig:stress}
\end{center}
\end{figure}

\fi

\end{document}